\documentclass[runningheads]{llncs}
\usepackage[colorlinks=true, citecolor=blue]{hyperref}
\usepackage{enumerate}
\usepackage{mathtools, amssymb, latexsym, amsmath, amsfonts, color}
\usepackage{array, fancyhdr, bm, listings}
\usepackage{algorithm,comment}
\usepackage{algorithmic}
\usepackage{subcaption}

\usepackage[title,toc,titletoc,header]{appendix}

\usepackage[title,toc,titletoc,header]{appendix}



\begin{document}
\title{Oblivious Set-maxima for Intersection of Convex Polygons}
\titlerunning{Oblivious Set-maxima Problem}
\author{Avah Banerjee\inst{1}\and
Dana Richards\inst{2}}
\authorrunning{A. Banerjee et al.}
%
\institute{Missouri S\&T, Rolla MO 6540, USA
\\ \email{banerjeeav@mst.edu}\and
George Mason University, Fairfax VA 22030, USA\\
\email{richards@gmu.edu}}
\maketitle              
\begin{abstract}
In this paper we revisit the well known set-maxima problem in the oblivious setting. Let $X=\{x_1,\ldots, x_n\}$ be a set of $n$ elements with an underlying total order. 
Let $\mathcal{S}=\{S_1,\ldots,S_m\}$ be a collection of $m$ distinct subsets of $X$.
The set-maxima problem asks to determine the maxima of all the sets in the collection. In the comparison tree model we are interested in determining the number of comparisons necessary and sufficient to solve the problem. 
We present an oblivious algorithm based on the lattice structure of the input set system. Our algorithm is simple and yet for many set systems gives a non-trivial improvement over known deterministic algorithms. We apply  our algorithm to a special  $\cal S$ which is determined by an intersection structure of convex polygons and show that $O(n)$ comparisons suffice. 
\end{abstract}
%
%
\section{Introduction}
The \textit{set-maxima} problem was first introduced by \cite{graham1980information}, in the context of finding lower bounds for shortest path problems. It was shown at the time that the decision tree bound is weak. 
The general problem remains important and unsolved. 

We define the problem. 
Let $X=\{x_1,\ldots, x_n\}$ be a set of $n$ elements with an underlying total order. 
Let $\mathcal{S}=\{S_1,\ldots,S_m\}$ be a collection of $m$ distinct subsets of $X$.
The set-maxima problem asks to determine the maxima of all the sets in the collection. Specifically we are interested in determining the number of comparisons necessary and sufficient to solve the problem. 
(We assume each element of $X$ occurs in at least two sets, since if it occurs in exactly one set we can postprocess it in constant time.)
In this model we assume that determining set memberships, computing the union / intersections of the sets etc. are free. 
We shall use the term {\em comparison complexity} to indicate that we are only dealing with the number of comparisons between elements and use the term {\em total complexity} to indicate the overall run time.
(To be clear, each $S_i$ will actually be a subset of $[1\ldots n]$, integer indices into $X$, so 
an implementation would not involve comparisons in set operations.)

The best known lower bound for the problem under the comparison tree model is no better than the trivial bound of $O(m + n)$. 
This was proved in \cite{graham1980information} using the $s$-\textit{uniqueness} property. 
The best upper bound for the problem is a combination of two upper bounds and is summarized as $O(\min(n \log n, n +  m(\min(2^{m}-1, n))))$. 
The $n \log n$ term comes from the following simple observation: if we sort the set $X$ then without any further comparisons we can determine the maximum of each set by simply scan the sorted list while doing membership queries. 
The second term is the results of the following procedure: for each element add it to the bucket (create one if it does not exists) representing the intersection of sets the element belongs to.
We have to create at most $n$ buckets since they are mutually disjoint.
Determine the maximum for each bucket, doing so takes at most $n$ comparisons.
Next for each of the $m$ sets determine the collection of at most $\min(n, 2^m-1)$ buckets the set has a non-empty intersection.
Compute the maxima of these buckets.
The second algorithms is only considered when $m=o(\log n)$.

\subsection{Previous and Related Work}
In  \cite{komlos1984linear} authors proposed an algorithm for the special set maxima problem
motivated by Graham et al. 
For the minimum spanning tree verification problem, $X$ is the set of weighted edges in a tree and the collection $\mathcal{S}$ consists of subsets of edges that join two non-adjacent vertices in the tree. 
Komlos' algorithm arbitrarily roots the tree and makes paths into pairs of paths to a common ancestor.
The algorithm makes $O(n \log((m+n)/m))$ comparisons.
In \cite{bar1992linear} authors gave the first general algorithm.
Their  ``rank-sequence algorithm'' determines a rank sequence $R$  according to the application domain. 
Specifically a rank sequence is an ordered sequence of $k$ ranks 
$n \ge r_k \ge \ldots \ge r_1 \ge 1$. 
The corresponding partition of $X$ is computed. 
Each $S_j$ is reduced to just those elements in one block of the partition.
When the elements are points and the sets are hyperplanes that form a projective space, 
it can be computed with linear comparisons, for a suitable rank sequence.
However, the rank-sequence algorithm is no better than the trivial algorithm above in the worst case. 
It was shown by  \cite{desper1994set}  that for some collection of subsets there are no good rank sequence for which the number of comparisons made by the algorithm is linear.
In \cite{liberatore1998matroid} Liberatore showed that this can be generalized using weighted matroids. 
One of the canonical examples of matroids is the graphic matroid. 
Generalized to binary matroids (since graphic matroids are also regular) this has been termed by Liberatore as the fundamental path maxima problem over such matroids. 
A cographic matroid is a dual of a graphic matroid. 
For a cographic matroid the problem can be solved in $O((m+n)\log^* n)$ (\cite{tarjan1982sensitivity}) comparisons.  
Liberatore generalized these results to a restricted class of matroids that can be constructed via direct-sums and 2-sums  and gave a $O(\min((m+n)\log^*(m+n), n\log n))$-comparison algorithm.
 
In the randomized setting the problem was fully solved by Goddard et. al \cite{goddard1993optimal} who proposed a sampling strategy to  based on the rank sequence algorithm.
They show that the expected number of comparisons  in their algorithm is 
$O(n \log {((m+n)/n)})$ which is optimal with respect to the comparison tree complexity. 
Their randomized algorithm can solve a more general problem of computing the largest $t$ elements for each subset $S_i$.

\subsection{Summary of Our Results}
In section \ref{sec: greedy} we give a greedy algorithm based on the overlapping subset structure of a lattice generated from the subset system $(X, \mathcal{S})$. Our algorithm is oblivious and hence can be implemented in a privacy sensitive environment. In section \ref{sec: range} we adapt our greedy algorithm to a special $\cal S$ which is determined by an intersection structure of convex polygons. 

\section{An Oblivious Algorithm}\label{sec: greedy}

There has only been one algorithm for the general set-maxima problem (\cite{bar1992linear}) and our algorithm is incomparable to that.
We concentrate on the underlying structure from the viewpoint of $X$.
Consider the bipartite graph $(S,X,E)$, $E=\{(i,j) | x_i\in S_j\}$.
Let $T_i=\{j | x_i\in S_j\}$.
Let $p= \sum_{j = 1}^{m}{|S_j|}= \sum_{i = 1}^{n}{|T_i|}$.
The {\em naive algorithm} makes $p-m$ comparisons; each $x_i$ offers itself to each $S_j$ in its $T_i$.
The first offer to each $S_j$ is accepted, without comparisons.
We improve on this algorithm by noting how the various $T_i$'s intersect. First we present a greedy strategy that is not oblivious. Letter we make it oblivious through simplification. The simplification does not effect the worst case complexity.

\subsection{A greedy (non-oblivious) algorithm}

When $T_i$ and $T_j$ intersect, $I_{i,j}= T_i \cap T_j \ne \emptyset$, comparing $x_i$ and $x_j$ helps.
If $x_i>x_j$ then we update $T_j$ to become $T_j\setminus I_{i,j}$, leaving the answer to the set-maxima problem unchanged.
(Of course, for each $k\in I_{i,j}, S_k$ becomes $S_k\setminus \{x_j\}$; however our algorithm is given in terms of the $T_i$'s.) 
When $x_i<x_j$ it is handled symmetrically.

Our greedy algorithm is: choose $i$ and $j$ so that $I_{i,j}$ is as large as possible, compare $x_i$ and $x_j$, update $T_i$ or $T_j$ as a result, and iterate.
We break ties in favor an $i$ and $j$ when $T_i=T_j$.
When no $T_i$ and $T_j$ intersect then we are essentially done.
The naive algorithm can finish up with no further comparisons.

Essentially the algorithm uses comparisons to successively change the input until the naive algorithm has no work to do.
We start with the initial input where the naive algorithm will make $p-m$ comparisons and after one comparison we create a new smaller version of the same problem, the ``induced problem.''
The induced problem has a smaller $p'= \sum_{i = 1}^{n}{|T_i|}= p-|I_{i,j}|$.
Using the naive algorithm to finish off the new problem, 
our situation improved from $p-m$ to $1+ p'-m= p-m+1-|I_{i,j}|$.
Since we maximize $|I_{i,j}|$ we greedily reduce the input for the next iteration.
Note that even when $|I_{i,j}|=1$, while the number of comparisons to finish stays the same, we move closer to the termination condition (where there are no intersections).

We need to improve our definitions to discuss the analysis.
Consider the following example, Figure 1, where we show the initial bipartite graph setting.
From the various $T_i$ we construct the (mixed) graphical diagram $G$.
The set of vertices is $X$.
A directed edge from $x_i$ to $x_j$ indicates $T_j \subset T_i$.
A double arrow indicates $T_i=T_j$.
An undirected edge indicates the remaining case when $|I_{i,j}|>0$.
The rightmost diagrams will be discussed in the next section.
There is no edge if $I_{i,j}=\emptyset$

\begin{figure}[h]
	\includegraphics[width=5in]{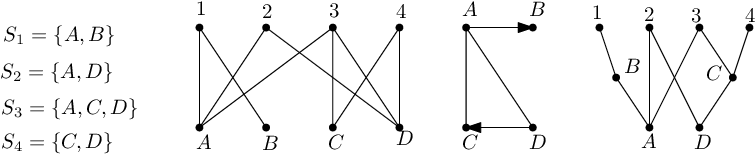}
	\centering
	\caption{The input, the bipartite setting, the intersection dependencies.}
\label{fig:ex1}
\end{figure}
\begin{figure}[h]
	\includegraphics[width=4.25in]{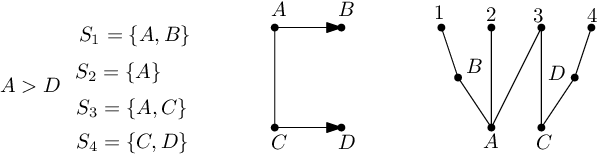}
	\centering
	\caption{One result of comparing $A$ and $D$.}
\label{fig:ex2}
\end{figure}
\begin{figure}[h]
	\includegraphics[width=4.25in]{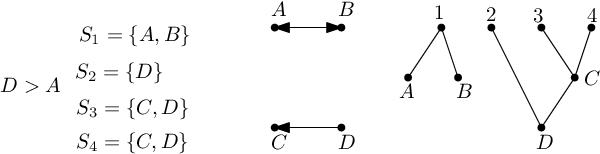}
	\centering
	\caption{Another result of comparing $A$ and $D$.}
\label{fig:ex3}
\end{figure}

Any time when $T_i=T_j$ we simply compare $x_i$ and $x_j$ and one or the other is eliminated, inasmuch as $T_i$ or $T_j$ will become empty. Clearly such comparisons occur $O(n)$ times.
For each directed edges between $x_i$ and $x_j$ a comparison will be made.
And each undirected edge a comparison will be made.
In all three cases the edge in our graphical diagram will disappear because the basic operation in the algorithm insures $T_i$ and $T_j$ will become disjoint.
Since each $T_i$ only loses elements no new edges will ever arise.
However existing edges can change type, as the Figures 2 and 3 shows.

The directed edges induce a directed acyclic subgraph. 
A directed path corresponds to a series of transitive subset containments.
The greedy algorithm will automatically work upwards through such chains, since the sizes of the intersections decreases.
Consider the case of $T_i\subset T_j \subset T_k$ and the result of comparing $x_j$ and $x_k$.
If $x_k>x_j$ then $x_j$ is eliminated and the result leaves $T_i \subset T_k$ and a subsequent comparison of $x_i$ and $x_k$; if $x_k<x_j$ then $T_k$ becomes disjoint from $T_j$ (as well as from $T_i$) and a subsequent comparison of $x_i$ and $x_j$.
Only two comparisons resolve the three containments, first a comparison against $x_j$ and then against $x_k$.
The algorithm is essentially using the \textit{transitive reduction} of the induced DAG; $x_i$ is not compared with both $x_j$ and $x_k$.
As a result, \textit{for the purpose of counting comparisons}, the greedy algorithm will have one comparison for each directed edge of the transitive reduction, i.e., with any transitively-induced directed edges removed.

So an upper bound on the number of comparisons made is the number of edges in the diagram, is $A+B+C$, where 
$$A=|\{(i,j) \:|\: T_i=T_j\}|,$$
$$B=|\{(i,j) \:|\: T_i\not\subseteq T_j\mbox{ and } T_j\not\subseteq T_i\mbox{ and } I_{i,j} \ne \emptyset \} |,$$
and $C=\sum_{i=1}^{n}|C_i|$, the number of edges in the transitive reduction, where
$$C_i=|\{ j \:|\: T_j\subset T_i \mbox{ and no } k, T_j\subset T_k\subset T_i \}|.$$
We call $C_i$ the {\em cover} of $x_i$.
Note that every comparison reduces the number $A+B+C$.

Because of the changing values of $T_i$ this can be a loose bound
for two reasons. 
First even without comparing $i$ and $j$ the intersection
can go away since another comparison involving, say, $i$ might shrink $T_i$.
Second, if we learn $x_i<x_j$ and later that $x_j<x_k$ then an edge between $i$ and $k$ can be processed without a comparison. Moreover this algorithm is not oblivious due to the above reasons. In the proceeding we consider a simplification which also happens to be oblivious.

\subsection{A simplified oblivious algorithm}\label{sec: simple}


The simplified algorithm, in the main loop, only considers comparing $x_i$ and $x_j$ 
when $T_i \subset T_j$, $T_j\subset T_i$ or $T_i=T_j$.
These correspond to the directed and doubly directed edges of the graph diagram discussed above.
As above, only the transitive reduction of the directed edges are used.
As a result when the main loop terminates there can still be intersecting covers.
It finishes with the naive algorithm.

Solely for the analysis, we assume a preprocessing step so each set $S_i$ is given an element $b_i$ that is only found in that set.
We add each $b_i$ to $X$ creating $X'$; so $n'=|X'|=m+n$.
This will not add any comparisons since the algorithm will not actually need to involve these new keys in any comparisons, since we can assume they all have a key value less than all the original keys.
The effect will be to have the graphical diagram have a vertex that represents each $S_j$;
therefore the analysis is more uniform.
We will also assume that any instances of $T_i=T_j$ are dealt with during preprocessing or as soon as they arise.
Let $G'$ be the new graphical representation.
The vertex set is the expanded $X'$ and there are only directed edges; bi-directed edges are processed immediately and undirected edges are ignored.

The rightmost diagrams in the figures illustrates the simplified algorithm.
Here the number $i$ does not represents the set $S_i$ but instead the unique element $b_i$.
Upward edges (no arrowheads shown) represents the subset containments.
Since it is the transitive reduction, each node points to its cover.
Because of the greedy nature of the algorithm the lower edges will be evaluated first, i.e., bottom-up.
As with the earlier analysis, each edge of the diagram corresponds to a comparisons that will be done by the simplified algorithm.
(The diagram can also be interpreted as a sublattice of the set containment lattice, where each $x_i$ is identified with its initial $T_i$.)

So the number of comparisons is $O(n+C')$, where $C'$ is the same as $C$ but is computed on the preprocessed graph $G'$. (That is, $C'$ is the number of edges in the rightmost diagrams.)
There are $O(n)$ comparisons for doubly directed edges (which are dealt with immediately and do not show in the diagram), and $C'$ for the directed edges.
Two observations.
First, undirected edges in the graphical diagram are ignored since the simplified algorithm does not make use of nonproper intersections.
Second, by including the $b_i$'s, the final comparisons of the naive
algorithm now become directed edges and are handled automatically.
It is easily seen that the simplified algorithms makes as many comparisons as the greedy algorithm.

It follows from these last observations that comparisons made by the simplified algorithm are known in advance.
In other words this approach is an oblivious algorithm. 
This has implications for the privacy aware computing domain, that are not discussed herein.

\begin{remark}
What is missing from the discussion so far is the effect of transitivity.
Not subset containment but the transitivity that is discovered amongst the elements of $X$.
There are two posets on $X$: subset and less-than.
The fact these are unrelated is what has made the set-maxima problem so difficult to solve over the years.
There are two response we can give.
First, we can however incorporate less-than transitivity in our original greedy algorithm.
We will not only maximize $|I_{i,j}|$.
To choose $i$ and $j$, for $x_i$ we compute the size of the intersection of $T_i$ and $T_j$ but if $x_i>x_j$, by transitivity, we should consider the size the intersection of $T_i$ and any $T_j \cup T_k$ where it is known $x_j>x_k$; we do the same for $x_i<x_j$.
We can greedily choose $i$ and $j$ to maximize the guaranteed cumulative intersection sizes.
Of course after performing the comparison all relevant $T_i$'s are updated.
Second, our greedy algorithm can look each connected components of the mixed graph, ignoring the directions of the edges.
Suppose a component with $n$ vertices has $m$ edges.
If $m$ is large then we could just sort those $n$ vertices and all the $m$ edges will be resolved without additional comparisons.
So the number of comparisons would be $O(\min \{ n+C', n\log n\})$.
So the worst case would also $O(n\log n)$.
We have not analyzed these approaches further.
Note that by judicious (not necessarily greedy) choices transitivity-based approaches might have a $o(n\log n)$ bound.
Indeed, this indicates the interplay between the two posets.
\end{remark}
\begin{remark}
It is not hard to find set systems where $C'$ is large. Suppose $G$ (left most graph in Figure \ref{fig:ex1} is the hypercube graph with $2n$ vertices (where $n=2^k$, for some number $k$). Then we see that $C' = O(n \log n)$. The hypercube graph poses a challenge for any algorithm trying to achieve a $o(n \log n)$ comparison complexity.
It is not a highly expanding graph (see for example \cite{alon1986eigenvalues} for a precise definition) and hence the rank-sequence algorithm cannot be used in this case.
However, it has $\Theta(n \log n)$ edges and hence comparing all the edges is cost prohibitive. Thus we believe that any attempt to solve the general set-maxima problem will require a better understanding of the combinatorial structure of the hypercube.
\end{remark}

\section{A Set-system Based On Intersections of Convex Polygons}\label{sec: range}
\begin{figure}[h]
	\includegraphics[width=5in]{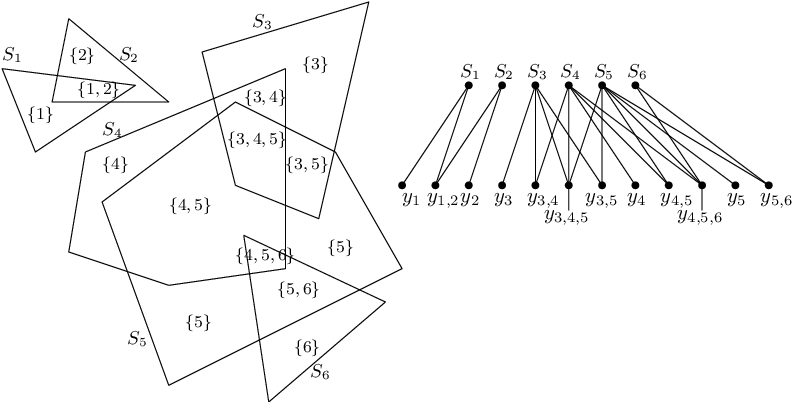}
	\centering
	\caption{A set-system from the intersection of six convex polygons $\{S_1,\ldots,S_6\}$. The set $X$ and the memberships are shown in right.}
\label{fig:convex}
\end{figure}
In this section we give an example of a set-system which naturally arises in certain geometric applications (such as image and shape processing).
Here we introduce and work on the planar case and we leave the higher dimensional case as an open problem.

Let ${\cal P} = \{P_1,\ldots,P_m\}$ be a set of $m$ convex polygons on the plane.
For any non-empty index set $I \subset [m]$, let $P_I = \bigcap_{j \in I}P_j$ be the intersection of all $P_j$'s where $j \in I$ (we take $P_{j} = P_{\{j\}}$). Let 
$${\cal J}_I = \{J \subset[m]\setminus \emptyset \mid I \subsetneq J\ \mbox{and}\ P_J \ne \emptyset\}$$ 
be the set of non-empty proper subsets of $I$. Note that ${\cal J}_I = \emptyset$  if $I$ is a singleton.
As an example (see Figure \ref{fig:convex}) take $I = \{4,5\}$. Then ${\cal J}_{\{4,5\}} = \{\{3,4,5\},\{4,5,6\}\}$.
If $P_I \setminus \cup_{J \in {\cal J}}P_J \ne \emptyset$ then we associate with the (not necessarily convex or connected) region an element $y_{I}$. Let $X = \{y_I\mid P_I \setminus \cup_{J \in {\cal J}}P_J \ne \emptyset\}$
be the set of all such elements. Further each element of $X$ has a key value which induces an unknown total order on $X$ ($|X| = n$ as before). We associate a set $S_i$ with the polygon $P_i$ where $S_i = \{y_I\in X\mid i \in I\}$. 
Let ${\cal S} = \{S_1,\ldots,S_m\}$.
Then the pair $(X, {\cal S})$ is the input to our set-maxima problem. According to the construction $m \le n$.


Additionally our problem has a parameter $k$ and we assume that each $P_j$ has at most $k$ sides.
Note that this does not restrict the cardinality of $S_j$. 
The algorithm from the previous section can be used by ignoring the geometry and converting it back into a set problem; we will use the simplified algorithm since the analysis is easier here. (The analysis only depends on the number of proper containments.)
We show that we can solve the set-maxima on the above set system with $O(n)$ comparisons when $k$ is fixed.

The algorithmic framework is the same as for the simplified algorithm.
Recall we will use a lattice (this is the lattice from the previous section) to speak of the graphical diagram.
Each vertex is associated with an $y_I$ ($x_i$ previously) and that is associated with a cover $T_I$.
In what follows we will identify $S_i$ with the corresponding convex polygon $P_i$.
The nodes in the first layer (in the reduced lattice) will be associated with each $P_i$; this can be regarded as the fictitious element $b_i$ discussed earlier. That is we assume for all $i$, $b_i = y_{\{i\}} \in X$ (again this is only for the purpose of our analysis).
We define a \emph{cover} for $y_I$ (as well as $P_I$) analogously.
$$C_I = \{J \mid J \subset I\ \mbox{and there are no}\ K \ \mbox{s.t.}\ J \subsetneq K \subsetneq I\ \mbox{and}\ y_K \in X\}$$
If $J$ is in the cover of $y_I$ then $P_I \subsetneq P_J$ and there exists no $P_K$ such that $P_I \subsetneq P_K \subsetneq P_J$ and alternatively $T_J \subsetneq T_K \subsetneq T_I$. We say $P_J$ is a \emph{cover} of $P_I$. Although covers are defined twice their intended meaning should be clear from the context. If $P_J$ is a cover of $P_I$ then there is a common edge between the polygons $P_I$ and $P_J$. 
At this point we could try to upper bound $C'$. However we make the following observation which leads us to work with a slightly different quantity $C''$. 
We call  $C'_I$ a \emph{complete sub-cover} of $y_I$ if the following holds: 1) $C'_I \subset C_I$ and 2) $I \subset \cup_{J \in C'_I}J$. If $C'_I$ only satisfies the latter condition we say $C'_I$ is a \emph{canopy} of $I$.
Note that (this is true even for the general case) we only need to compare $y_I$ with $y_J$'s where $J \in C_I'$. Since by doing so we end up comparing $y_I$ to some element of each of the sets it belongs to. Let $C'' = \sum_{y_I}|C'_I|$ for some collection of complete sub-cover of the elements. In what follows we show that  $C'' =  O(n)$.   




%

\begin{figure}[h]
	\includegraphics[width=4cm]{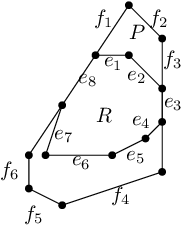}
	\centering
	\caption{Polygonal chains.}
\label{fig:chains}
\end{figure}

\noindent{\em Moving forward, for readability, we will refer to our $P_I$s  simply as polygons $P$, $R$, $S$, etc.}

Assume $P$ is a cover $R$. 
Let $c_{P,R}$ be a polygonal chain of successive edges of $R$ which are not (part of) some edge of $P$. 
In  Figure \ref{fig:chains} we see two such polygonal chains.
An upper chain $c_1 = (e_1, e_2)$ and a lower chain $c_2 = (e_4,e_5,e_6,e_7)$.
Let $C_{P,R}$ be the set of all such chains formed by the intersection of $P$ and $R$.
Note that $e_8 \subset f_1$ and $e_3 \subset f_3$ are not part of any chain.
If we treat a chain as a set consisting of edges, then we can define the  set operators on a pair of such chains.
From the remarks above we have $|C_{P,R}|>0$.

\begin{figure}[h]
	\includegraphics[width=5cm]{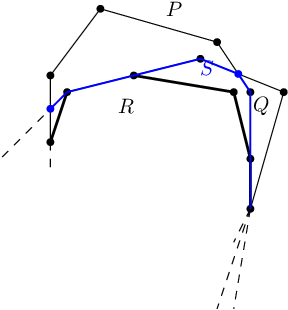}
	\centering
	\caption{A case where $P$ and $Q$ are not covers of $R$.}
\label{fig:pqr}
\end{figure}

\noindent{\bf Observation 1.}
If $P, Q$ are in the cover of $R$ then for any chain $c \in C_{P,R}$ and $c' \in C_{Q,R}$ $c \cap c' = \emptyset$. 

\begin{proof}
Let there be some $c \in C_{P,R}$ and $c' \in C_{Q,R}$ such that $c \cap c' \neq \emptyset$.
Then from Figure \ref{fig:pqr} we see that there is a region $S$ with positive area. Hence we have $R \subsetneq S \subsetneq P$ and $R \subsetneq S \subsetneq Q$ which contradicts our assumption that $P, Q$ are covers of $R$.
\end{proof}

\noindent{\bf Observation 2.}
If $T$ is a set of $l$ elements and $S$ be collection of subsets from $T$ each of which has size $\ge l-k$, then every set in $S$ contains some element from any subset of size $\ge k+1$ of $T$. 

\begin{lemma}\label{lmm: main geo}
If each $P_i$ has at most $k$ sides then for the set system described above $|C_I'| \le k+1$.
\end{lemma}

\begin{proof}
Let $P_I$ be a  region formed by the intersection of $r$ polygons, $|T_I|=r \ge 1$. Suppose $P_I$ has $l$-sides.
Let these polygons be $\{P_1,\ldots,P_r\}$.
For each  polygon $P_i$,  at least $l-k$ of the sides of $P_I$ will be part of some polygonal chain (i.e., in $C_{P_i,P_I}$) of $P_i$.
This can be easily seen since $P_i$ has at most $k$ sides and is convex, at most $k$ edges of $P_I$ can also be part of the edges of $P_i$ (in the degenerate case $I = \{i\}$).
Let $D_i$ be the collection of all edges of $P_I$ that are in some chain of $P_i$.
And let $D=\{D_1,D_2,\ldots,D_r\}$ be the collection of all these sets.
From observation 2 we can pick a set $W$ of  $k+1$ edges of $P_I$ such that every polygon $P_i$ has at least one edge  from the set $C_{P_i,P_I}$.
Let
\begin{align}\label{eq: canopy}
    C_I' = \{J_e, e\in W\mid\ \mbox{ $\forall j \in J_e$}\ e\in D_j\}=\{J_1,\ldots,J_{k+1}\}.
\end{align}

Clearly $C_I'$ is a canopy (that is $I \subset_{J \in C_I'}J$). 
If $C_I'$ is also a sub-cover then we are done. 
Otherwise,  there exists  $J$ and $J'$ in $C_I'$ such that replacing $J$ and $J'$ with $J \cup J'$ produces a canopy of smaller size.
In general if $C_I'$ is not a sub-cover then there exists $S \subset C_I'$ such that $C_I'\setminus S \cup (\cup_{J \in S}J)$  is also a canopy.
We iterate until $C_I'$ becomes a sub-cover.
It follows then that the ultimate size of $C_I'$ is $\le k+1$.
\end{proof}

\begin{remark}
From equation \ref{eq: canopy} we already have $|C'_I| \le k+1$. However it may be the case that for some $S \subset C'_I$, there are no $y_J \in X$ for all $J \in S$. That is these $J$'s correspond to empty intersections. Thus including them will increase the size of $X$ in a non-trivial way would contribute to an increase in the number of comparisons. Hence we specifically look for a complete sub-cover.
\end{remark}


\begin{theorem}
For the set-maxima problem arsing from intersection of polygons of bounded number of sides there is a oblivious algorithm which uses $O(n)$ comparisons.
\end{theorem}

\begin{proof}
From lemma \ref{lmm: main geo} we see that $C'' = \sum_{y_I}|C'_I| \le n (k+1) = O(n)$. Combining this with our observations from section \ref{sec: simple} the result follows.
\end{proof}


\ifx false
Parameterization of the results in terms of $k$ is important. 
Without any restriction on the polygons, it is possible to represent any arbitrary set system in this geometric setting.
The construction is easy.
Take all the points of $X$ to be on a circle.
Then any subset of points are the corners of a convex polygon.
Restricting $k$ allows the geometry to play a role.
Removing the constraint of being a full problem and achieving the bound seems difficult.
Without it the size of the covers can exceed $k$; amortization does not seem to help.
Another approach keeps the ``fullness" but populates some regions with dummy points not in $X$. Unfortunately
the number of dummy points can be quadratic in $m$,
without a guaranteed improvement in the number of comparisons.
These observations even apply to axis-oriented rectangles,
but with more hope of success.
\fi

\begin{remark} (Total Complexity)
Total complexity depends on how the input is given to us.
Since the above algorithm is oblivious the construction of the complete sub-covers and hence the lattice is done once given a fixed set-system ($\cal S$) and remains unchanged for different total orderings of $X$. Once the lattice is computed solving the set-maxima using the oblivious algorithm  takes $O(n)$ time in total.
So if the input is given to us in the form of the lattice then total complexity is the same as the comparison complexity. 
On the other hand if only the polygons are given (as ordered sequence of points) then we need to construct the lattice.
We argue that constructing the lattice takes $\Tilde{O}(m^4k^2)$\footnote{This is the soft-O notation.} total time in the RAM model. This follows from that fact that the intersection of a  set of at most $k$ sided $m$ polygons has at most $O(km^2)$ points of intersections. 
We iteratively process each polygon: start with an arbitrary pair of polygons and process their intersection. When adding the next polygon determine the new intersections (if any) it generates. This takes $\Tilde{O}(m^3k^2)$ per polygon (see for example \cite{muller1978finding}). Since there are $m$ polygons the observations follows.
\end{remark}

\begin{remark}(Extension to Polytopes)
We believe a similar result holds when we replace the polygons with their $d$-dimensional counterparts ($d \ge 3$) where $d$ is fixed and the number of facets of the polytopes are bounded. However we have not worked through this case and leave this as an open problem. 
\end{remark}


  \bibliographystyle{elsarticle-num} 
  \bibliography{mybibfile}

\begin{thebibliography}{1}
\expandafter\ifx\csname url\endcsname\relax
  \def\url#1{\texttt{#1}}\fi
\expandafter\ifx\csname urlprefix\endcsname\relax\def\urlprefix{URL }\fi
\expandafter\ifx\csname href\endcsname\relax
  \def\href#1#2{#2} \def\path#1{#1}\fi

\bibitem{graham1980information}
R.~L. Graham, A.~C. Yao, F.~F. Yao, Information bounds are weak in the shortest
  distance problem, Journal of the ACM (JACM) 27~(3) (1980) 428--444.

\bibitem{komlos1984linear}
J.~N. Koml{\'o}s, Linear verification for spanning trees, in: Foundations of
  Computer Science, 1984. 25th Annual Symposium on, IEEE, 1984, pp. 201--206.

\bibitem{bar1992linear}
A.~Bar-Noy, R.~Motwan, J.~Naor, A linear time approach to the set maxima
  problem, SIAM Journal on Discrete Mathematics 5~(1) (1992) 1--9.

\bibitem{desper1994set}
R.~Desper, The set-maxima problem: an overview, Master's thesis, Rutgers
  University (1994).

\bibitem{liberatore1998matroid}
V.~Liberatore, Matroid decomposition methods for the set maxima problem, in:
  SODA, 1998, pp. 400--409.

\bibitem{tarjan1982sensitivity}
R.~E. Tarjan, Sensitivity analysis of minimum spanning trees and shortest path
  trees, Information Processing Letters 14~(1) (1982) 30--33.

\bibitem{goddard1993optimal}
W.~Goddard, C.~Kenyon, V.~King, L.~J. Schulman, Optimal randomized algorithms
  for local sorting and set-maxima, SIAM Journal on Computing 22~(2) (1993)
  272--283.

\bibitem{alon1986eigenvalues}
N.~Alon, Eigenvalues, geometric expanders, sorting in rounds, and ramsey
  theory, Combinatorica 6~(3) (1986) 207--219.

\bibitem{muller1978finding}
D.~E. Muller, F.~P. Preparata, Finding the intersection of two convex
  polyhedra, Theoretical Computer Science 7~(2) (1978) 217--236.

\end{thebibliography}


\end{document}